# Muon spin rotation and neutron scattering investigations of the B-site ordered double perovskite $Sr_2DyRuO_6$

D.T. Adroja[1,2*], Shivani Sharma[1,6&], C. Ritter[3£], A.D. Hillier[1], C.V. Tomy[4], R. Singh[5], R. I. Smith[1], M. Koza[3], A. Sundaresan[6], S. Langridge[1]

[1]*ISIS facility, Rutherford Appleton Laboratory, Chilton Oxon, OX11 0QX*
[2]*Highly Correlated Matter Research Group, Physics Department, University of Johannesburg, Auckland Park 2006, South Africa*
[3]*Institut Laue-Langevin, 71 Avenue des Martyrs, CS 20156, 38042, Grenoble Cedex 9, France*
[4]*Department of Physics, Indian Institute of Technology Bombay, Mumbai 400 076, India*
[5]*Indian Institutes of Science Education and Research, Bhopal, India*
[6]*Jawaharlal Nehru Centre for Advanced Scientific Research, Jakkur, Bangalore, India*
(date: 29-07-2019)

The magnetic ground state of B-site ordered double perovskite $Sr_2DyRuO_6$ has been investigated using muon spin rotation and relaxation (μSR), neutron powder diffraction (NPD) and inelastic neutron scattering (INS), in addition to heat capacity and magnetic susceptibility (*ac* and *dc*) measurements. A clear signature of a long-range ordered magnetic ground state has been observed in the heat capacity data, which exhibit two sharp anomalies at 39.5 and 36 K found as well in the magnetic data. Further confirmation of long-range magnetic ordering comes from a sharp drop in the muon initial asymmetry and a peak in the relaxation rate at 40 K, along with a weak anomaly near 36 K. Based on temperature dependent NPD, the low temperature magnetic structure contains two interpenetrating lattices of $Dy^{3+}$ and $Ru^{5+}$, forming an antiferromagnetic ground state below 39.5 K with magnetic propagation vector $k = (0,0,0)$. The magnetic moments of $Dy^{3+}$ and $Ru^{5+}$ at 3.5 K are pointing along the crystallographic *b*-axis with values of $\mu^{Dy} = 4.92(10)$ $\mu_B$ and $\mu^{Ru} = 1.94(7)$ $\mu_B$, respectively. The temperature dependence of the $Ru^{5+}$ moments follows a mean field type behaviour, while that of the $Dy^{3+}$ moments exhibits a deviation indicating that the primary magnetic ordering is induced by the order of the 4*d*-electrons of $Ru^{5+}$ rather than that of its proper 4*f*- $Dy^{3+}$ electrons. The origin of the second anomaly observed in the heat capacity data at 36.5 K must be connected to a very small spin reorientation as the NPD studies do not reveal any clear change in the observed magnetic Bragg peaks' positions or intensities between these two transitions. INS measurements reveal the presence of crystal field excitations (CEF) in the paramagnetic state with overall CEF splitting of 73.8 meV, in agreement with the point change model calculations, and spin wave excitations below 9 meV at 7 K. Above $T_N$, the spin wave excitations transform into a broad diffuse scattering indicating the presence of short-range dynamic magnetic correlations.

E-mail: *Devashibhai.adroja@stfc.ac.uk, &shivani.sharma@stfc.ac.uk, £ritter@ill.fr

# I. Introduction

Geometrically frustrated antiferromagnetic (AFM) materials have attracted considerable interest over the past few years, motivated by their tendency to form rather exotic magnetic ground states such as the spin-glass, spin-liquid, or spin-ice instead of long-range magnetic order in apparent defiance of the third law of thermodynamics [1,2,3,4,5,6]. Among the four "canonical" geometrically frustrated lattices: triangular planar, kagome, pyrochlore, and face-centered cubic (fcc), the latter has recently gained strong attention [6,7,8,9]. In real materials, the fcc magnetic lattice is conveniently realized in the B-site ordered double perovskites, $A_2BB'O_6$ [10]. Here a magnetic ion resides on the B'-site, while B can be either magnetic or non-magnetic and A-site is non-magnetic. Both the B and B' sites constitute interpenetrating face-centered cubic sublattices in which, if the exchange constraints between nearest neighbours are AFM, the basic criteria for geometric frustration are satisfied [6,9,11,12,13,14,15].

Recently, the double perovskites compounds with general formula $A_2BB'O_6$, with A alkaline metals, B rare earth metals and B' transition metal, have attracted considerable attention due to their interesting physical properties as well as possible applications in renewable energy and spintronic devices [6, 10,11,12,13,14,15,16,17]. Within this class of materials, there are compounds with properties such as a high Curie temperature, $T_C$ [18,19], phase separation, [20] a high magnetoresistance, [21,22] a metal- insulator transition [23,24], and half-metallic antiferromagnets [21,25]. Besides the interesting fundamental physics, double perovskite materials are important for optoelectronic applications and technology [26].This huge variety of properties has its origin in the possibility of doping and substituting the perovskite structure at the A- and B-sites, allowing tailoring of the electronic, crystal, and magnetic structure of the compounds, which, in turn, interact with each other. $Sr_2FeMoO_6$ was the first double perovskite for which a high magnetoresistance at room temperature was reported ($T_C \sim 420$ K) [27]. By electron doping in similar compounds, the Curie temperatures rises to 635 K for $Sr_2CrReO_6$ [28,29,30] and 750 K for $Sr_2CrOsO_6$ [31] which is so far the highest $T_C$ observed in ferrimagnetic double perovskites. A special type of double exchange interactions [32,33] was shown to be responsible for the high magnetic transition temperatures and the strong spin polarization in double perovskites where B and B' cations are in a mixed valence state [34]. Adoption of integer valences leads to reduced $T_C$ or to antiferromagnetic order [34,35].

Among the antiferromagnetically ordered double perovskites, $Sr_2LnRuO_6$ (Ln = rare earth, Y, Ho, Yb and Lu etc.) compounds exhibit many interesting properties, for example the presence of two magnetic phase transitions and strong geometrical frustration above the magnetic ordering up to as high as 300 K, confirmed via heat capacity and inelastic neutron scattering measurements, respectively [36,37,38,39,40]. Recent neutron diffraction studies on $Sr_2YRuO_6$ reveal that at the first transition temperature only half of the Ru-layers order magnetically while the other half (alternatively) reveals short range ordering and below the second phase transition the system exhibits a type-I AFM ground state [37]. Although the presence of frustration has been observed in many double perovskites compounds, its origin is not clear at present. In addition, diffuse scattering has been observed in the compounds $Sr_2YRuO_6$, $La_2NaRuO_6$, $La_2NaOsO_6$ and $Sr_2YbRuO_6$ [37,38,41] in which $La_2NaRuO_6$ reveals a single magnetic transition below 15 K to an incommensurate magnetic ground state whereas $La_2NaOsO_6$ does not exhibit any long range order down to 4 K on the quasi-FCC lattices as a result of geometrical frustration [11,41]. These results motivate the investigation of other double perovskites compounds in order to understand the presence of geometrical frustration and its effect on the magnetic ground state. We have therefore studied the detailed dynamic and static magnetic properties of Sr$_2$DyRuO$_6$ (SDRO) using magnetization, heat capacity, muon spin resonance/rotation (μSR), neutron powder diffraction (NPD) and inelastic neutron scattering (INS) measurements. SDRO exhibits a magnetic anomaly ~ 40 K, which is suspected to be associated with the long range ordering [42,43]. An exchange bias effect in SDRO has also been observed below AFM ordering and the possible cause for the observed effect was suspected to be linked to Dzyaloshinskii-Moriya (DM) interactions present in this geometrically frustrated system [42]. DFT results report that the main contribution to the spin moment comes from the *f*- orbitals, with a considerable role of the *d*-orbitals and suggest that SDRO will behave as a conductor and semiconductor for spin-up and spin-down orientations, respectively [44]. However, no direct evidence or studies about the electronic or magnetic structure/ground state is available on SDRO in the existing literature. The present work will fill the gap to understand the low temperature magnetic behaviour of SDRO and provide an ideal example to compare with the available data of other geometrically frustrated double perovskite having two magnetic cations at the B and B' sites.

## II. Experimental details

The polycrystalline sample (10 gm) of Sr$_2$DyRuO$_6$ (SDRO) was prepared by solid-state reaction from stoichiometric amounts of SrCO$_3$, RuO$_2$ and Dy$_2$O$_3$ (Aldrich 99.99 %) which were mixed in an agate mortar pestle and pressed into pellets. These pellets were then annealed at 1123 K for 12 h and sintered at 1253 K for 24 h, with frequent regrinding and repelletizing. The structure characterization at 300 K was carried out using the GEM time-of-flight (TOF) neutron powder diffractometer (NPD) at the ISIS neutron Facility, UK. The *dc*-magnetic susceptibility and magnetization isotherm were measured using a SQUID magnetometer (Quantum Design, MPMS). Heat capacity measurements were performed using a relaxation technique by a commercial system (Quantum Design, PPMS) in the temperature range of 1.8–100 K. The *ac*-susceptibility was measured using the same Quantum design's PPMS. To investigate the magnetic structure/ground state, low temperature NPD measurements were performed using the constant wavelength (λ = 2.396 Å) high intensity diffractometer D20 between 1.7 and 50 K at ILL Grenoble, France. High-resolution data were recorded as well at the ILL on the powder diffractometer D2B using λ = 1.594 and 2.399 Å. All the diffraction data have been analyzed using the Rietveld refinement program Fullprof [45]. The μSR experiments were carried out using the MuSR spectrometer in the longitudinal geometry at the ISIS muon facility, UK. We have performed zero-field (ZF) and longitudinal-field (LF) μSR measurements between 1.5 and 50 K and LF field between 0 and 2500 G. The powder sample (thickness ~2 mm) was mounted onto a 99.995+% pure silver plate using GE-varnish and were covered with 18 micron silver foil. Inelastic neutron scattering measurements were performed on the time-of-flight spectrometers MERLIN at ISIS Facility and IN6 at ILL, Grenoble. We use a powder sample of SDRO in an annular Al-can of outer diameter 40 mm on MERLIN and 20 mm on IN6.

## III. Results and discussion

### (a) Room temperature structural characterization

Figure 1 shows the NPD pattern of SDRO collected at 300 K from the 34.96° detector bank of the GEM diffractometer. The structure was refined using the monoclinic space group *P*2$_1$/*n*, assuming a 1:1 ordering of the Dy$^{3+}$ and Ru$^{5+}$ cations. The Dy$^{3+}$ and Ru$^{5+}$ cations occupy distinct Wyckoff sites, 2*c* and 2*d*, respectively, resulting in the ordered arrangement. No impurity peaks were detected within the resolution limit. The refined lattice parameters at 300 K are *a* = 5.7774(2) Å, *b* = 5.7948(2) Å, *c* = 8.1848(2) Å, *β*= 90.181(3)° and V = 276.88(1) Å$^3$. The determined lattice and structural parameters are in good agreement with the existing literature [42,43,44]. The refinement did not give any evidence for a possible site-disorder between the $Dy^{3+}$ and $Ru^{5+}$ cations. Therefore, our results confirm the ordered double perovskite structure of SDRO.

**(b) Physical properties:**

Figures 2 shows the measured heat capacity of SDRO as a function of temperature for zero-field (ZF) and in applied fields of 1 and 9 Tesla (T). Two anomalies are evident at 39.6 K and 36.5 K (see the inset in Fig.2) in the ZF heat capacity data, which disappear in a field of 9 T. The anomalies are more clearly visible in the first order derivative which is presented as an inset in the same figure. At 1T field, the lower transition does not change much, but the higher transition broadens and moves towards higher temperature. Similar two anomalies/transitions were also reported for isostructural $Sr_2LnRuO_6$ (Ln=Y, Ho, Yb and Tb) and identified as antiferromagnetic ordering temperatures ($T_{N2}$ and $T_{N1}$). The anomalies were situated at 24 and 29 K for Y [36,37], 36 and 40 K for Yb [38], 32 and 26 K for Lu [46], and 15 and 36 K for Ho-based systems [47]. On the other hand, the heat capacity study on the cubic $Ba_2DyRuO_6$ reveals only a single anomaly at 47 K [48], similar to $La_2NaRuO_6$ [11,41]. Further, $Sr_2FeOsO_6$ exhibits two magnetic transitions at $T_{N1}$=140 K and $T_{N2}$=67 K, where both the Fe and Os moments order and the second transition is associated with the change in magnetic structure from AF1 to AF2 [49]. Considering the observation of a spin gap only below $T_{N2}$ in the inelastic study of $Sr_2FeOsO_6$, it was suggested that spin-orbit coupling is important for ground state selection in this compound [49]. This suggests that the two anomalies observed in the heat capacity of SDRO are possibly associated either to the separate long range magnetic ordering of the Ru and Dy moments or to a spin reorientation transition.

The temperature-dependence of the *dc*-magnetic susceptibilities ($\chi_{dc}$) of SDRO in various applied magnetic fields is shown in Fig. 3 in zero-field cooled (ZFC). The increase of susceptibility below 42 K, irrespective of the applied field value indicates the emergence of long range magnetic ordering. With further decreasing temperature, $\chi_{dc}$ first increases and exhibits a sharp peak near 40 K for B = 25 Oe, nearly matching the first anomaly observed in the heat capacity data (39.6 K). For B = 500 Oe, the peak in the susceptibility becomes quite broad exhibiting a plateau. For B = 1000 Oe, there is no visible peak and the susceptibility keeps on increasing down to 2 K. As it is difficult to identify the magnetic ordering temperature directly from the $\chi_{dc}$ behaviour, the first order derivative of $\chi_{dc}$ is plotted in the left inset of the same figure and shows for all three field values a clear peak at 40 K. This is in accordance with the heat capacity results where the first anomaly was observed at $T_{N1}$ = 39.5 K. No direct signature of a second anomaly as found in the heat capacity data ($T_{N2}$ = 36.5 K) is evident from the susceptibility data (Fig. 3) for B = 500 Oe and 1000 Oe while the derivative points to $T_{N2}$ = 36 K for B = 25 Oe (inset of Fig. 3). Only an indirect indication of $T_{N2}$ can be found by the rate of change of $\chi_{dc}$ for the B = 500 and 1000 Oe field curves which changes below 35 K. Black arrows are pointing in Fig. 3 to the changes in slopes which are presented by the dashed black lines. As a function of the applied field strength, the values of $T_{N2}$ can be estimated as 36.7, 34 and 32 K for B = 25, 500 and 1000 Oe, respectively, The Curie-Weiss fit of the inverse susceptibility for 500 Oe data is also shown in the same figure in the right inset. The estimated total value of the effective paramagnetic moment is 10.42 $\mu_B$ which is slightly smaller than the theoretical value arising from the combined paramagnetic contribution of $Dy^{3+}$ and $Ru^{5+}$ ions which amounts to 11.33 $\mu_B$ $\left(\mu_{eff.} = \sqrt{\left(\mu_{eff.}^{Dy3+}\right)^2 + \left(\mu_{eff.}^{Ru5+}\right)^2}\right)$.

The isothermal magnetization behaviour (*M-vs-H*) of SDRO is presented in Fig. 4 as a function of applied magnetic field (H) at selected temperatures ranging from 2 to 50 K. To perform these measurements, the sample was cooled each time from the paramagnetic state (300 K) to avoid any magnetic history effect. The *M-vs-H* isotherm at 50 K is almost linear in H, as expected for a paramagnet state. At 2 K, the *M-vs-H* curve initially increases rapidly with increasing field up to 0.4 T before it exhibits an almost linear field dependence. The rapid increase in magnetization is also observed for temperatures between 2 and 40 K, but the value above which the field dependence is showing a linear behaviour is reducing with increasing temperature. A hysteresis is observed when cycling the field (see inset of Fig. 3) with the overall magnetization and the width of the hysteresis decreasing with increasing temperature. The observed weak ferromagnetic-type behaviour in *M-vs-H* data at low fields in the antiferromagnetic state, which is also observed in other $Sr_2LnRuO_6$ (R= rare earth) compounds [38,39,40], has been attributed to the contribution of the weak ferromagnetic component from the DM interaction. The magnetization attains a value of only ~ 4.31 $\mu_B$ at 7 T, which is very small compared to the theoretical value of the saturation magnetization of $Dy^{3+}$ ion ($g_L J = 10$ $\mu_B$, where $g_L = 1.243$ is Landé g-factor of $Dy^{3+}$), but in good agreement with existing literature [42]. The reduction of the saturation value of the magnetization compared to the full moment of $Dy^{3+}$ and the additional contribution expected from $Ru^{5+}$ is attributed to the effect of the crystal field (CEF). The CEF will split the J=15/2 ground multiplet of $Dy^{3+}$ into 8 doublets in the paramagnetic state and 16 singlets in the magnetic ordered states due to Zeeman splitting. Hence, the ordered state moment value will reduce and will depend on the ground state wave functions. This is also supported by our inelastic neutron scattering study discussed in Section-(e).

Figure 5 represents the real $\chi'_{ac}$ and the imaginary part $\chi''_{ac}$ of the temperature dependent *ac* susceptibility of SDRO at frequencies ranging from 10 Hz to 10 kHz. Both anomalies as observed in the heat capacity and $\chi_{dc}$ are visible in the real and imaginary parts. The anomaly at $T_{N1} = 40$ K is revealed by a sharp jump in both $\chi'_{ac}$ and $\chi''_{ac}$, and is frequency independent whereas the anomaly at $T_{N2}$ creates a broad shoulder to the main transition and shows a weak frequency dispersion. This indicates that the second ordering is possibly associated with a very small change in the spin structure near $T_{N2}$.

**(c) μSR measurements**

In order to gain insight as to whether the two observed phase transitions in the heat capacity originate from the magnetic ordering, we have investigated SDRO using the μSR technique. μSR is a local microscopic probe and sensitive to extremely small internal fields and ideal to detect spatially inhomogeneous magnetic features and is extensively applied to investigate small change in magnetism [50]. It is interesting to mention that μSR studies on double perovskites have provided important information on the magnetic ground state of these systems [41,51,52,53], including the information of microscopic co-existence of magnetic and non-magnetic phases in $Ba_2PrRu_{9.9}Ir_{0.1}O_6$ [54]. For the present study, zero-field (ZF) and longitudinal field (at constant temperature) μSR measurements have been performed. Figure 6 shows the muon initial asymmetry versus time spectra at several temperatures between 1.2 and 90 K measured in ZF. The analysis of μSR spectra was carried out using stretched exponential function with constant background.

$$G_z(t) = A_0 * \exp(-(\lambda t)^\beta) + A_{bg} \qquad (1)$$

Here $A_0$ is the muon initial asymmetry, $\lambda$ the muon relaxation rate. If the exponent $\beta = 2$, the function becomes Gaussian while for $\beta = 1$, the function becomes Lorentzian. $A_{bg}$ is the constant background arising from muons stopping on the Ag-sample holder. Further $\beta < 1$, describes inhomogeneous dynamic relaxation where the relaxation is locally exponential but the local rates are distributed [55]. It has no basic theoretical justification, but is often used as a convenient characterization of an a priori unknown distribution of relaxation rates. As in SDRO we would expect different internal fields for muons stopping on the DyO$_6$ octahedral site and on the RuO$_6$ octahedral site, the present approach to fit data with the use of a stretched exponential function therefore seems appropriate.

Fig. 7(b) shows that at high temperature (i.e. above the Néel temperature) the µSR spectra exhibit a moderate relaxation rate, which is due to the spin fluctuation from the Ru$^{5+}$ and Dy$^{3+}$ moments. With decreasing temperature, the relaxation rate increases and exhibits a peak near 40 K, followed by a rapid loss of muon initial asymmetry (Fig. 7a) below 40 K. Between 1.2 and 40 K, the asymmetry loss is almost 2/3. In the polycrystalline sample, we expect three components of internal field at the muon stopping sites. The loss of 2/3 can be understood as arising from implanted muons that do not see the component of internal field that is parallel to the incident muon beam (1/3 component), while the remaining two components (2/3 component, transverse components, which give oscillations) of the internal field will be seen by the muons. As the internal field from the Dy$^{3+}$ and Ru$^{5+}$ moments are expected to be larger and inhomogeneous, the oscillations in the muon time evolution asymmetry will be damped rapidly. Therefore, due to the pulsed width of the ISIS muon beam, it is difficult to observe these oscillations in 2/3 component as this signal damps very fast in short time window close to zero time. We therefore attribute the observed jump in $A_0$ at 40 K to long-range magnetic ordering of both Dy$^{3+}$ and Ru$^{5+}$ moments, as also observed in the heat capacity and magnetization measurements. This is also in agreement with our finding from the neutron diffraction study discussed below. Furthermore, if we look in detail at the behaviour of $\lambda(T)$ near 36 K, then there is weak evidence of a second phase transition in $\lambda$(T). However, $A_0$ does not reveal any sign of a second transition, which one would expect, as the system is in a complete long range magnetic order state below 40 K and hence cannot lose further asymmetry. We therefore attribute the weak change in the $\lambda$(T) near 36 K due to a very small change in the spin configuration, which was clearly detected in the heat capacity. Furthermore, $\beta$(T) reveals temperature dependent behaviour. $\beta$(T) starts to decrease with temperature from ~1 at 90 K to ~0.7 at 40 K and then exhibits a sharp jump to 1 below $T_{N1}$ and remains the same to lower temperature. It would be interesting to note that very recent µSR study on La$_2$NaRuO$_6$ and La$_2$NaOsO$_6$ [41] exhibits similar features in $\lambda(T)$, but only exhibits a single magnetic transition. The µSR spectra of these compounds were fitted better to stretched exponent function. Furthermore, there are no clear signs of frequency oscillations. This indicates that internal fields at the muon stopping sites, which are most likely close to the Oxygen ions due to the negative charge on O$^{2-}$ and positive charge on muon are larger than 800 G. The width limit of ISIS muon pulse does not allow collecting data at very low time close to zero. The ordered moment of Dy$^{3+}$ is quite high compared to Ru$^{5+}$ (discussed in neutron diffraction section), and it is highly possible that muons stopping in DyO$_6$ octahedra will see a higher internal field than those stopping in RuO$_6$ octahedra. We have also performed LF-measurements at 5, 35 and 50 K up to 2.5 kG field (see Fig. S1 of the supplementary material (SM) [56]). At 5 and 35 K the initial asymmetry gradually recovers to 0.12 and 0.15 respectively at 2.5 kG field. Furthermore, the relaxation rate also exhibits a gradual decrease with applied field. On the other hand at 50 K, the initial asymmetry does not change much with applied field, but the relaxation rate decreases slowly with applied field.

**(d) Neutron powder diffraction studies**

To explore the origin of the two anomalies observed in various experiments, temperature dependent NPD measurements have been performed on the high intensity diffractometer D20 between 1.7 and 50 K with λ = 2.399 Å (Fig. 8). Long scans of 45 minutes have been performed at 50, 40, 30, 20, 10 and 1.7 K and shorter scans of ~11 min were recorded between these temperatures while raising the temperature with a constant ramp of 1 K/5 mins. Figure 8(a) represents the NPD patterns at 1.7 and 50 K, along with the difference curve plotted in green. Several new peaks emerge and some existing ones increase in intensity at low temperature indicating the onset of long-range magnetic ordering. The magnetic reflections can be indexed with $h+k+l$ being odd. Figure 8(b) is the 3D plot of the temperature dependent D20 data in a limited 2θ range in which the magnetic peaks are indexed using the nuclear unit cell, i.e. the magnetic propagation vector $k$ = (0, 0, 0). This is consistent with A-type antiferromagnetic order where ferromagnetic sheets within the *ab*-plane are antiferromagnetically stacked along the long *c*-axis with both the magnetic and crystal unit cell of same size. The presence of the (001) reflection indicates that the magnetic moments have components in the *ab*-plane. The strongest magnetic reflection represents in fact two reflections which can be indexed as (010)/(100), which due to the pseudosymmetry where the *a~b* can't be resolved. Magnetic symmetry analysis for the two possible magnetic sites of $Dy^{3+}$ on 2c and $Ru^{5+}$ on 2d was done in space group $P2_1/n$ with $k$ = (0, 0, 0) using the program BASIREPS [57,58]. Both sites possess the same two allowed irreducible representations (IR) with each having three basis vectors (BV) (Table II). The two IRs are differing in the directions of the ferro- (F) or antiferromagnetic (AF) couplings between the symmetry related sites within the unit cell. While IR1 allows an AF coupling along the unit cell *b*-direction with F couplings allowed along *a*- and *c*-directions. The IR2 describes just the opposite with a F coupling along *b* and AF couplings along *a* and *c*. Testing both IRs, it becomes clear that due to the above mentioned pseudosymmetry, it is possible to refine the low temperature data assuming either an AF coupling along *a* (IR1) or along *b* (IR2). In both cases a single BV is sufficient to refine the magnetic structure, there is no clear indication of a ferromagnetic contribution to the magnetic scattering. As even our high resolution data from D2B at 3 K using λ = 1.594 Å are not able to differentiate between the two models (Fig. 9b), additional NPD data have been collected at 3.5 K as well on the high-resolution powder diffractometer D2B using the longer wavelength of λ = 2.399 Å in order to determine whether the moments are pointing in the *a*- or *b*- directions (Fig.10). The only magnetic peak which allows to differentiate between these two possibilities is the 100/010 doublet. Figure 10 shows the Rietveld refined NPD pattern measured at 3.5 K. The insert shows the enlarged view of the (010)/ (100) peaks. It clearly shows that there is magnetic intensity at the 100 reflection but not on the 010 reflection, which confirms that the magnetic moments are pointing in the *b*-direction. The resultant magnetic structure is shown in Fig. 11 in which the $Ru^{5+}$ moments are represented in green color and the $Dy^{3+}$ moments in red color. Using this model, the magnetic structure was refined using the high resolution data collected on D2B at 3 K and λ = 1.595 Å, with the magnetic intensity modelled as a second phase containing only the Dy and Ru atoms. The magnetic form factor used for $Ru^{5+}$ is the one determined empirically in [59]. Figure 9(b) shows this refinement where the lower red set of tick marks correspond to the magnetic Bragg reflections. Table I shows the relevant bond lengths and angles at T = 3 K together with those determined at T = 50 K (Fig.

9(a)): Cooling through the magnetic transition leads to a compression of the RuO$_6$ octahedra whereas the DyO$_6$ octahedra elongate.

A cyclic structure refinement using the temperature dependent data from D20 allowed the determination of the thermal dependence of the magnetic moments of Dy$^{3+}$ and Ru$^{5+}$ and is shown in Fig. 12(a) with the normalized moments plotted in Fig. 12 (b). It can be seen that for the present system, the Ru$^{5+}$ moment attains saturation at a faster rate near ~20 K compared to Dy$^{3+}$ which attains saturation only well below ~10 K. This behaviour is similar to other members of this family, like Sr$_2$HoRuO$_6$ and Sr$_2$TbRuO$_6$ [39]. However, the direction of the magnetic moments of Dy$^{3+}$ and Ru$^{5+}$ are different in the present system from those of the Ho and Tb based double ruthenates. While both the rare earth and the Ru$^{5+}$ moments are along the *c*-axis in Sr$_2$HoRuO$_6$, they are canted by 20° from the *c*-axis for Sr$_2$TbRuO$_6$[39]. The moments of Ru and Pr in the *ac*-plane (i.e. tilted away from the c-axis) were also found in Ba$_2$PrRuO$_6$ [54]. Furthermore, the magnetic structure of Sr$_2$ErRuO$_6$ shows Ru$^{5+}$ and Er$^{3+}$ moments are mainly aligned along the *c*-axis of the structure, forming an angle of ~6° with the *c*-axis in the case of the Er$^{3+}$ sublattice and ~15° for the Ru$^{5+}$ moment [60]. In the present studied system, both the Dy$^{3+}$ and Ru$^{5+}$ moments are pointing along the *b*-axis. The values of the Dy$^{3+}$ and the Ru$^{5+}$ moments at 1.7 K are $\mu_{Dy3+}$ = 4.92(10) $\mu_B$ and $\mu_{Ru+5}$ = 1.94(7) $\mu_B$. The strong reduction of $\mu_{Dy3+}$ compared to the expected value of 10 $\mu_B$ is the line with similar discrepancies observed for the other rare earth Ru-based perovskites [39]. The Ru$^{5+}$ moment value is similar to those reported for other members of this family [37,39,47] and points to the fact that in these systems the Ru-O-O-Ru interactions are the strongest magnetic interactions and control the Ru ordering. The similar values of $\mu_{Ru5+}$ and of the magnetic transition temperatures in the different Sr$_2$*Ln*RuO$_6$ systems [39] are explained by the weakness of the *Ln*-O-Ru interactions. As exemplified by the very low magnetic transition temperature of Dy$_2$O$_3$ ($T_N$ = 1.2 K), Dy-O-Dy interactions are in general very weak. In the well-ordered double perovskite SDRO, only weaker super-super exchange Dy-O-O-Dy interactions are present which can't be the origin of the primary ordering of Dy$^{3+}$ at 39 K as supported by the order parameters given in Fig. 12(b). The temperature dependence of the Ru$^{5+}$ moments exhibits a mean field power law behaviour with a critical exponent β = 0.56(1) whereas that of the Dy$^{3+}$ moments deviates from the power law. Hence, it appears that Ru$^{5+}$ induces the rare earth ordering in these systems leading to a simultaneous ordering of Dy$^{3+}$ at the same temperature, as also previously reported for Sr$_2$LnRuO$_6$ (Ln = Ho and Tb) [39] as well as in R$_2$RuO$_5$ [61,62,63]. It should be noted that there appear neither appreciable changes in the magnetic peak profiles nor new magnetic Bragg peaks in the temperature dependent data across the second anomaly (~36.5 K). In particular we did not detect any additional superlattice peak in the temperature region between $T_{N1}$ and $T_{N2}$, such as the one created by a propagation vector k = (½, ½, 0) found by Bernardo et al. [36] in Sr$_2$YRuO$_6$. Furthermore there is no indication for short range correlations in the background below $T_{N1}$ = 40 K. This indicates that the second transition might be associated with very small changes in the spin structure/spin reorientation which are beyond the detection limit even of the high intensity data. Furthermore, the symmetry would allow a FM component on Dy and Ru moments, which cannot be better detected in the ND data, but might be responsible for weak ferromagnetic hysteresis observed in the magnetization isotherm at low temperature.

### (e) Inelastic neutron scattering studies

It is very important to understand the origin of reduced magnetic moment and strength of anisotropy in SDRO. We therefore have performed inelastic neutron scattering (INS) measurements on SDRO at various temperatures. Figure 13(a-d) shows the color contour maps of the scattering intensity, energy transfer vs momentum transfer (Q), at various temperatures between 7 K and 45 K with neutron incident energy, $E_i$ = 15 meV. At 7 K strong band of excitation can be seen near 3.25 meV, and weak scattering intensity near 5.63, 6.8 and 8.9 meV, which are more clear when presented in 1D intensity versus energy plot between Q=0 to 2 Å$^{-1}$ (Fig. 13e-f). At 30 K, the 3.25 meV excitation softens and it seems scattering intensity emerges out from Q = 1.1 Å$^{-1}$, which is the magnetic Bragg peak with index (0 1 0)/ (1 0 0). Further, increasing the temperature to 37 K, the inelastic scattering broadens and transforms into diffuse scattering with an energy width of ~8 meV. At 45 and 55 K (above $T_N$), we have seen the presence of weak and broad diffuse scattering in the elastic cut (see Fig. S2 in SM [56]), which suggests the presence of magnetic frustration/short-range correlations above $T_N$ in SDRO. As we did not see clear sign of diffuse scattering in the diffraction data on D20, it may suggest that the diffuse scattering has a dynamic nature. Further from the data at 5 K, we confirm the presence of spin wave at 7 K with spin gap of ~3.25 meV and zone boundary energy of 8.9 meV. It is to be noted that the observed scattering could be also partly interpreted as Zeeman splitting of the Dy low energy crystal field excitations (CEF) levels below $T_N$. Now we compare the value of spin gap (defined as a peak position in the energy cut, q-integrated close to AFM zone centre for the powder samples) observed in the present system with those reported with other double perovskites systems [64,65,66,67]. Our INS on $Sr_2YRuO_6$ reveals a spin gap of 5 meV [64]. The spin gap of 1.8(8) meV and 6(1) meV has been observed in $La_2LiRuO_6$ and $La_2LiOsO_6$, respectively [65], 2.57(4) meV in $La_2MgIrO_6$, 2.09(3) in $La_2ZnIrO_6$ [66], 5 meV in $Ba_2YRuO_6$ [67], 2.75 meV in $La_2NaRuO_6$ [41], 19 meV in $Sr_2ScOsO_6$ [68]. These results may suggest that the spin gap arising from the transition metal d-electrons having strong spin-orbital coupling.

Now we discuss the crystal field excitations measured using $E_i$ = 130 meV ($E_i$ = 250 meV data are given in the SM). Figure 14(a-b) shows the color contour maps of the scattering intensity, energy transfer versus momentum transfer, at 7 K and 100 K and Fig. 14(c-e) shows the Q-integrated energy cuts from low-Q, medium-Q and high-Q data. At 7 K and at low-Q, strong intensity of scattering is observed near 46.6 meV and 73.8 meV. Further, weak peak can be seen near 90.8 meV at higher-Q, but has a lower intensity at lower-Q. When we plotted the Q-dependence, energy-integrated intensity of these peaks (see Fig. S3 in SM [56]) the intensity of 46.6 and 73.8 meV decreases initially and starts to increase at higher Q, while that of 90.7 meV peak increases with Q. Furthermore the intensity of 90.7 meV follows $Q^2$ behaviour (see the inset in Fig. S3(b)) as expected for phonon scattering. These observations indicate that 46.6 and 73.8 meV peaks at low-Q are due to the crystal field excitations of $Dy^{3+}$ ion, while the 90.7 meV peak is due to purely phonon scattering. The increase of the intensity of 46.3 and 73.8 meV peak at higher-Q indicates that at nearly same position there are phonon modes. This might suggest the presence of phonon and CEF coupling as they have very similar energy scale. The assignment of CEF and phonon peaks seen in 130 meV data was also confirmed through the measurements with Ei=250 meV at 7 K and 120 K and data are plot in Fig. S4 of SM [56].

Now we look at the data of 130 meV at 100 K (250 meV at 120 K) and it is clear that a new strong peak near 37.3 meV appears at 100 K (same in 250 meV data at 120 K) and 73.8 meV peak has shifted to lower energy. It is also likely that 46.6 meV peak is also shifted to lower energy. This observed new peak near 37.3 meV we attribute due to the excited state transition from the CEF levels below 10 meV, which get populated at 100 K and gives this transition. The shift in the peaks could be due to various origins, i.e., magnetoelastic coupling and Zeeman field at 7 K.

We now discuss the CEF splitting of the $Dy^{3+}$ ($4f^9$) ions based on the CEF Hamiltonian in order to provide further support for our interpretation of the INS spectra. The point symmetry of the $Dy^{3+}$ ions is triclinic (1 or Ci) in the monoclinic $P2_1/c$ crystal structure of SDRO. In such a low symmetry, the CEF Hamiltonian requires 15 CEF parameters to be estimated from the INS spectra which is a difficult task. Considering an odd-number of electrons, $4f^9$ of $Dy^{3+}$ ions and Kramers' theorem [53], which says that for an odd number of electrons, minimum degeneracy of CEF levels should be two-fold (or doubly degenerate) in the paramagnetic state. We therefore expect J=15/2 ground state (16-fold degeneracy = 2J + 1) should split into 8 CEF levels with 2-fold degeneracy of each levels above $T_N$. Further, below the magnetic ordering these 8 doublets will split into 16 singlets. Hence if all CEF excitations from the ground state are allowed then in the paramagnetic state, we should expect a minimum of 7 CEF excitations from the ground state, in addition we will have additional CEF excitations due to the excited state transitions at $T$=100 K>$T_N$, if there are low energy CEF levels exist and the excited state transitions are allowed. Given that we have observed only two CEF transitions at 46.6 meV and 73.8 meV (in addition to 4 spin wave type excitations below 9 meV) at $T$ = 7 K and one additional excited state transition at 37.3 meV at 100 K a quantitative analysis of the INS data based on CEF model is not feasible. We have therefore used a point charge mode to estimate the 15 CEF parameters (including 11 complex parameters total 26 CEF parameters) of the CEF Hamiltonian and calculated INS spectra based on these estimated parameters. The simulated INS CEF spectrum at 7 K and is given in Fig. S5 of SM [56]. The simulation shows qualitative agreement with the experimental data giving overall CEF splitting 86.6 meV, which is in good agreement with observed CEF splitting of 73.8 meV.

## IV. Conclusions

Our combined heat capacity, magnetization, μSR, neutron diffraction and inelastic neutron scattering results demonstrate that $Sr_2DyRuO_6$ (SDRO) exhibits a long-range ordered magnetic ground state below 40 K. The heat capacity reveals a clear sign of two magnetic transitions, which are also indirectly supported through the magnetic susceptibility (both *ac* and *dc*) measurements. Our μSR and neutron diffraction studies further provide direct support of long-range magnetic ordering below 40 K. The neutron diffraction study shows that all observed magnetic Bragg peaks between 2 and 40 K can be indexed using the magnetic propagation vector k=(0, 0, 0). The magnetic structure shows that both the Dy and Ru atoms are arranged in type-I antiferromagnetic structure, which consists of interpenetrating sublattices of $Dy^{3+}$ and $Ru^{5+}$ atoms. In the *ab*-plane, the $Dy^{3+}$ and $Ru^{5+}$ moments are aligned AFM to each other, while along the *c*-axis they show FM coupling. Interestingly, the magnetic ordering is primarily governed by the 4*d*-moment on the $Ru^{5+}$ atoms and the $Dy^{3+}$ moments follows the Ru ordering at the same temperature ($T_{N1}$). In addition, it appears that the interactions responsible for the Dy ordering are weaker than the interactions responsible for the Ru ordering. The origin of the second anomaly in the heat capacity still remains an open question as the neutron diffraction study shows only one magnetic transition at 40 K and further single crystal neutron diffraction study will be important to understand the origin of two-phase transitions in SDRO. From the inelastic neutron scattering study, we have estimated the spin gap of 3.25 meV in the spin wave spectrum with maximum zone boundary energy of 8.9 meV. Furthermore, we have also discussed the presence of crystal field excitations and their role in the observed reduced moment of the $Dy^{3+}$ ions estimated through the neutron diffraction. The total CEF splitting observed in the experimental data agrees very well with that calculated using the point change model for the $Dy^{3+}$ ion. The present work will generate interest in condensed matter theory to develop a realistic model to find out a common origin of two magnetic phase transitions in double perovskite family.

**Acknowledgement:**

We would like to thank L. Pascut for his help during the D20 experiment, SS wants to thank India-Nanomission, DST and ISIS Facility for funding and DTA and ADH would like to thank CMPC-STFC, Grant No. CMPC- 515 09108, for financial support. We thank the ISIS facility for beam time for µSR and neutron measurements, RB1310039, DOI: 10.5286/ISIS.E.24090174.

**Table I**: Selected bond lengths (Å) and bond angles (°) across AFM ordering.

|           | 50 K       | 3.5 K      |           | 50 K       | 3.5 K      |
|-----------|------------|------------|-----------|------------|------------|
| **Ru-O1** | 1.938(7)   | 1.952(6)   | **O1-Ru-O2** | 88.8(3)  | 89.8(3)    |
| **Ru-O2** | 1.948(7)   | 1.965(6)   | **O1-Ru-O3** | 89.3(3)  | 90.6(3)    |
| **Ru-O3** | 1.966(6)   | 1.954(5)   | **O2-Ru-O3** | 89.1(3)  | 89.7(3)    |
| **Dy-O1** | 2.232(7)   | 2.225(6)   | **O1-Dy-O2** | 92.1(3)  | 91.7(3)    |
| **Dy-O2** | 2.250(7)   | 2.238(6)   | **O1-Dy-O3** | 91.5(3)  | 89.3(2)    |
| **Dy-O3** | 2.217(6)   | 2.219(5)   | **O2-Dy-O3** | 91.3(3)  | 88.3(2)    |
| **Ru-O1-Dy** | 158.2(4) | 157.3(4)  | **Ru-O3-Dy** | 155.5(4) | 156.3(3)   |
| **Ru-O2-Dy** | 154.4(4) | 155.0(4)  | **Ru-Dy**    | 4.0883(1) | 4.0832(1) |
|           |            |            |           | 4.0951(1)  | 4.0903(9)  |
| **Ru-Ru** | 5.7741(3)  | 5.7675(1)  | **Dy-Dy** | 5.7741(3)  | 5.7675(1)  |
|           | 5.7747(3)  | 5.7915(1)  |           | 5.7984(1)  | 5.7915(1)  |
|           | 5.8088(3)  | 5.8017(1)  |           | 5.8088(3)  | 5.8017(1)  |

**Table II:** Basis vectors (BV) of the allowed irreducible representations (IR) for $k = (0, 0, 0)$ for the Wykoff positions 2$c$ (Dy) and 2$e$ (Ru) of the space group $P2_1/n$.

| IR1 | BV1 | BV2 | BV3 |
|---|---|---|---|
| *x, y, z* | 100 | 010 | 001 |
| *-x+½, y+½, -z+½* | -100 | 010 | 00-1 |
| **IR2** | | | |
| *x, y, z* | 100 | 010 | 001 |
| *-x+½, y+½, -z+½* | 100 | 0-10 | 001 |

**Figure captions:**

**Fig. 1:** Rietveld fit to NPD patterns collected at 300 K on GEM diffractometer at ISIS: Black crosses show observed data points; red line shows calculated profile; lower blue line is difference profile (obs.-cal.); black vertical markers indicate Bragg peak positions.

**Fig. 2:** Heat capacity measurements in the low temperature range measured in zero-field and applied field of 1 and 9 T. The inset shows $dC_p/dT$ vs T near the magnetic ordering.

**Fig. 3:** (a) *dc* magnetic susceptibility ($\chi_{dc}$) measured at various applied magnetic field in zero-field. The dashed and dotted lines show the guide to eye. Inset (i) shows the first order derivation of ZFC $\chi_{dc}$ and the inset (ii) shows the inverse susceptibility ($1/\chi_{dc}$) measured in applied field of 500 Oe. The solid line shows the fit to Curie-Weiss behaviour.

**Fig. 4:** Magnetization isotherms measured at various temperatures ranging from 2 to 50 K. the insets shows the enlarge view at lower fields. A clear hysteresis can be seen for T < 40 K.

**Fig. 5:** (a) Real and (b) imaginary component of *ac*-susceptibility ($\chi_{ac}$) measured at various frequencies as a function of temperature.

**Fig. 6:** Zero-field μSR spectra measured at various temperatures. The experimental data are shown by the symbols and the solid line shows fit to the data using stretch exponent function.

**Fig. 7:** The temperature dependent fit parameters obtained from the Zero-Field μSR spectra. (a) Initial muon asymmetry versus temperature, (b) relaxation rate versus temperature and (c) exponent $\beta$ versus temperature. The dotted line in (a) shows the temperature independent back ground from the sample holder.

**Fig. 8:** Neutron diffraction pattern at 1.7 (blue curve) and 50 K (red curve) measured using D20. At 1.7 K, extra peaks are presented compare to 50 K, which are due to magnetic ordering of the $Dy^{3+}$ and $Ru^{5+}$ moments. The green line at the bottom represents the difference curve (1.7 K – 50 K) and shows only the magnetic Bragg peaks. (b) Thermal evolution of magnetic peak profiles between 2 and 50 K. The arrows show the magnetic Bragg peaks.

**Fig. 9:** Rietveld fit to NPD patterns collected at (a) 50 and (b) 3 K with $\lambda = 1.594$ Å on D2B diffractometer at ILL: Red circles show observed data points; black line is calculated profile; lower blue line is difference profile (obs.-cal.); upper green vertical markers indicate nuclear Bragg peak positions; lower red vertical markers in (b) indicate magnetic Bragg peak positions.

**Fig. 10:** Rietveld fit to NPD patterns collected at 3.5 K with $\lambda = 2.399$ Å on D2B diffractometer at ILL Red circles show observed data points; black line is calculated profile; lower blue line is difference profile (obs.-cal.); upper green vertical markers indicate nuclear Bragg peak positions; lower red vertical markers indicate magnetic Bragg peak positions. The inset enlarge the view near 24° to highlight the magnetic peak (010)/(100).

**Fig. 11:** The magnetic structure of $Sr_2DyRuO_6$ for $k = (0, 0, 0)$. The $Dy^{3+}$ and $Ru^{5+}$ moments are shown in green and red colors respectively.

**Fig. 12:** Thermal variation of (a) $Dy^{3+}$ and $Ru^{5+}$ moments and (b) Normalized moments of $Dy^{3+}$ and $Ru^{5+}$.

**Fig. 13:** (a-d) The color contour maps of scattering intensity versus momentum transfer at various temperature measure with an incident energy of $E_i = 15$ meV at various temperatures on MERLIN. (e-f) The Q-integrated (Q=0 to 2 Å$^{-1}$) energy cuts at various temperatures between 7 and 55 K.

**Fig. 14:** (a-b) The color couture maps of scattering intensity versus momentum transfer at various temperature measure with an incident energy of $E_i = 130$ meV at 7 and 100 K on MERLIN. (c-f) The Q-integrated energy cuts at low-Q, medium-Q and at high-Q at 7 and 100 K.

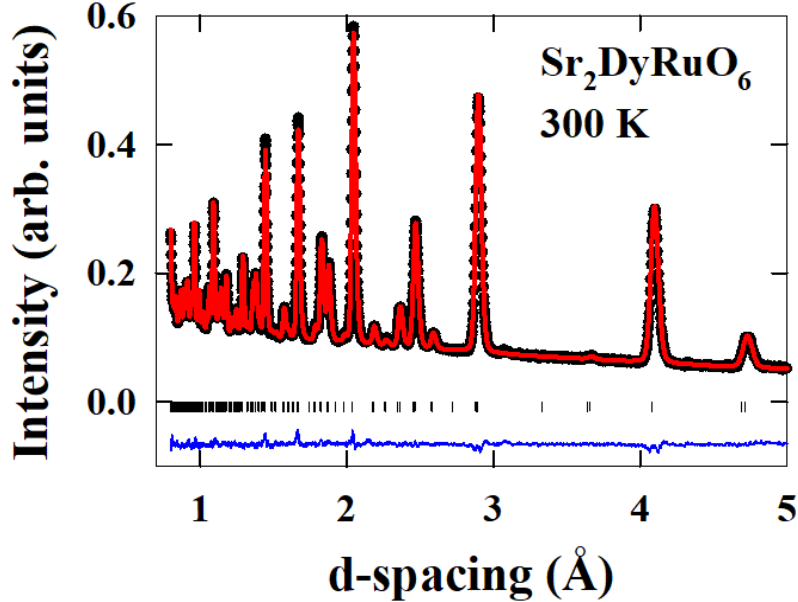

**Fig. 1:** Rietveld fit to NPD patterns collected at 300 K on GEM diffractometer at ISIS: Black crosses show observed data points; red line shows calculated profile; lower blue line is difference profile (obs.-cal.); black vertical markers indicate Bragg peak positions.

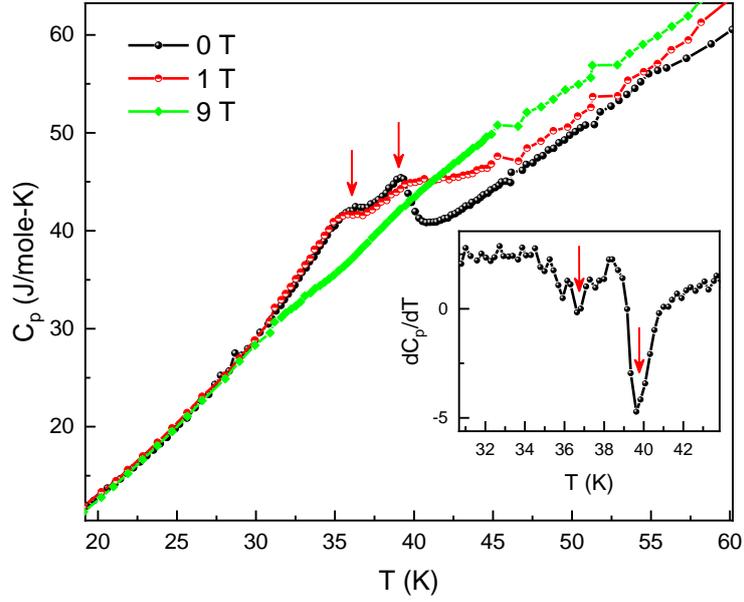

**Fig. 2:** Heat capacity measurements in the low temperature range measured in zero-field and applied field of 1 and 9 T. The inset shows $dC_p/dT$ vs T near the magnetic ordering.

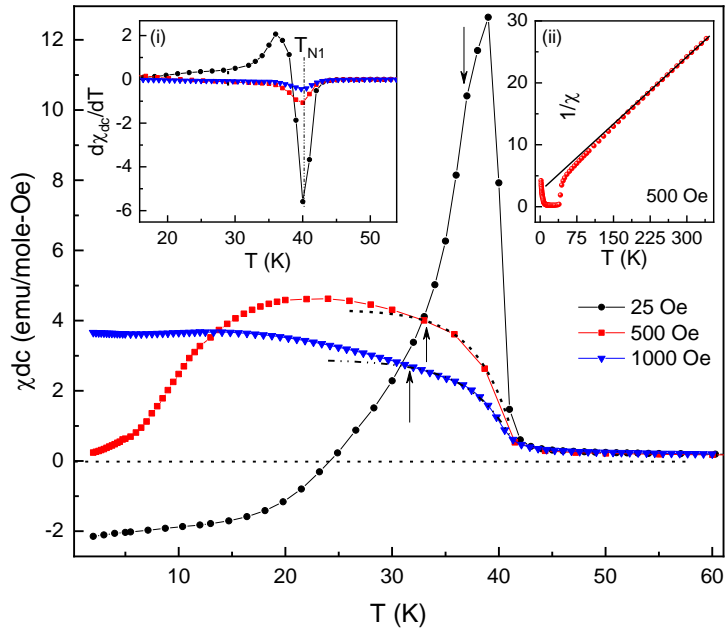

**Fig. 3:** (a) *dc* magnetic susceptibility ($\chi_{dc}$) measured at various applied magnetic field in zero-field. The dashed and dotted lines show the guide to eye. Inset (i) shows the first order derivation of ZFC $\chi_{dc}$ and the inset (ii) shows the inverse susceptibility ($1/\chi_{dc}$) measured in an applied field of 500 Oe. The solid line shows the fit to Curie-Weiss behavior. It is to be noted that the negative magnetization seen below 25 K in 25 Oe data is due to an artefact due to trapped field in the superconducting magnet of the SQUID magnetometer [69].

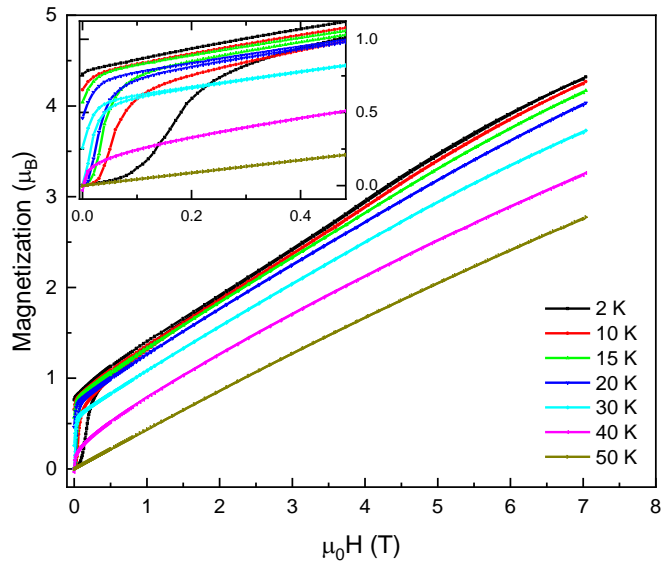

**Fig. 4:** Magnetization isotherms measured at various temperatures ranging from 2 to 50 K. the insets shows the enlarge view at lower fields. A clear hysteresis can be seen for T < 40 K.

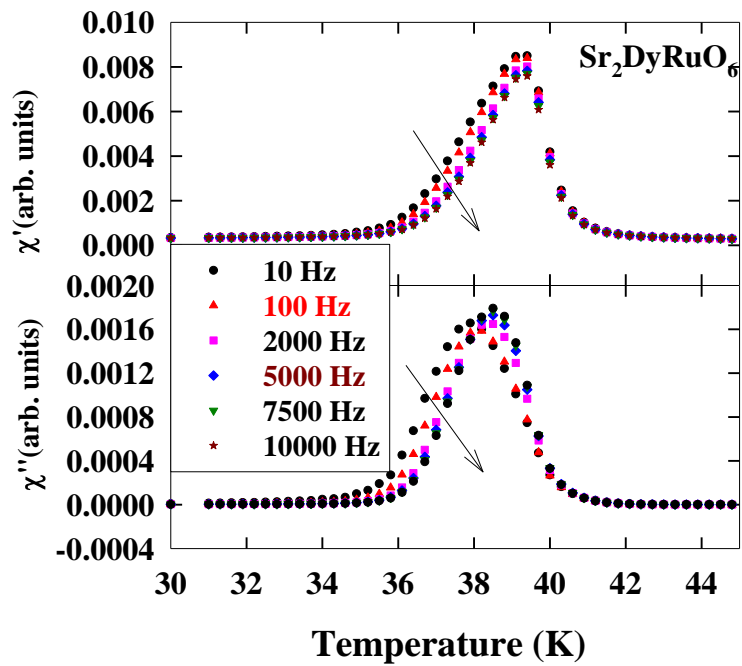

**Fig. 5:** (a) Real and (b) imaginary component of *ac*-susceptibility ($\chi_{ac}$) measured at various frequencies as a function of temperature.

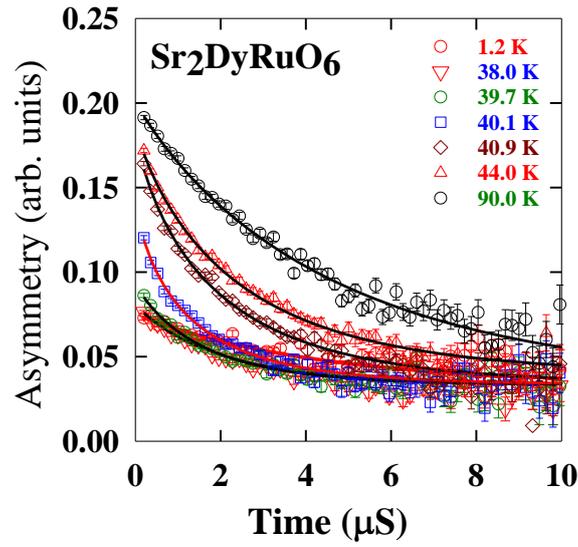

**Fig. 6:** Zero-field μSR spectra measured at various temperatures. The experimental data are shown by the symbols and the solid line shows fit to the data using stretch exponent function.

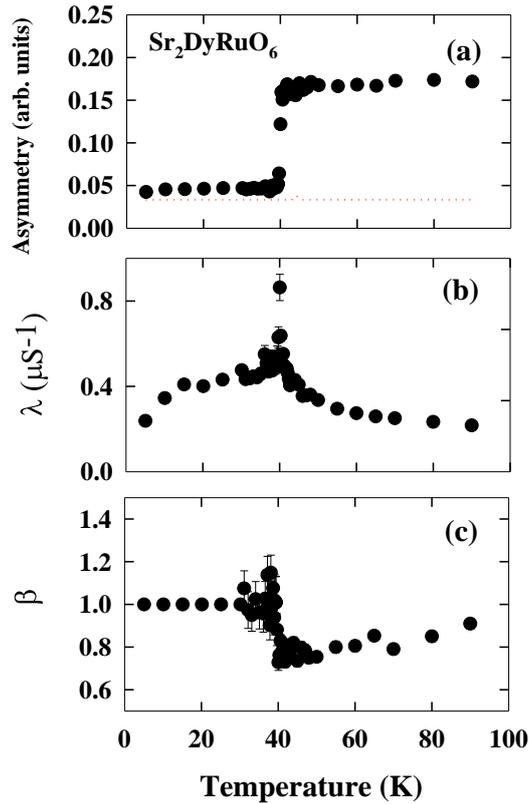

**Fig. 7:** The temperature dependent fit parameters obtained from the Zero-Field μSR spectra. (a) Initial muon asymmetry versus temperature, (b) relaxation rate versus temperature and (c) exponent $\beta$ versus temperature. The dotted line in (a) shows the temperature independent back ground from the sample holder.

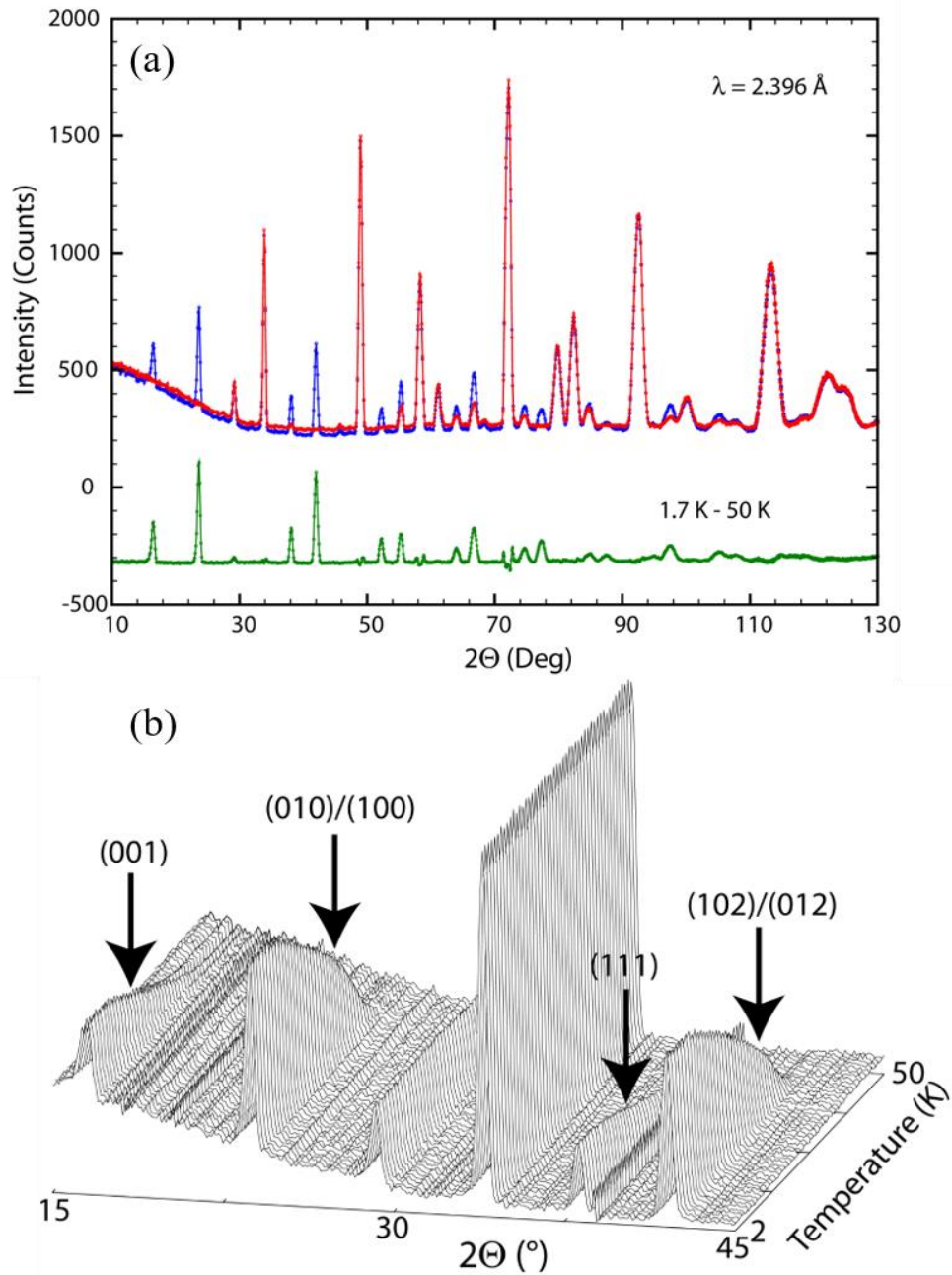

**Fig. 8:** Neutron diffraction pattern at 1.7 (blue curve) and 50 K (red curve) measured using D20. At 1.7 K, extra peaks are presented compare to 50 K, which are due to magnetic ordering of the $Dy^{3+}$ and $Ru^{5+}$ moments. The green line at the bottom represents the difference curve (1.7 K – 50 K) and shows only the magnetic Bragg peaks. (b) Thermal evolution of magnetic peak profiles between 2 and 50 K. The arrows show the magnetic Bragg peaks.

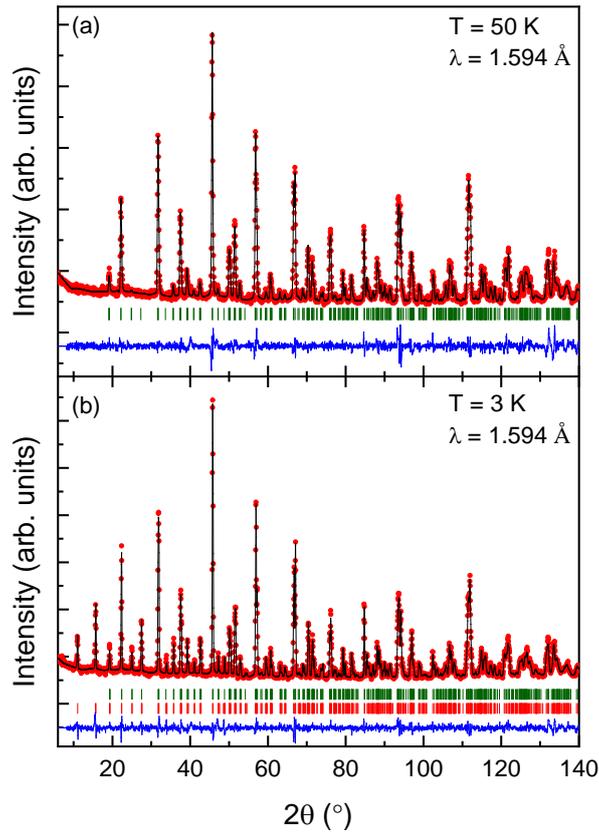

**Fig. 9:** Rietveld fit to NPD patterns collected at (a) 50 and (b) 3 K with λ = 1.594 Å on D2B diffractometer at ILL: Red circles show observed data points; black line is calculated profile; lower blue line is difference profile (obs.-cal.); upper green vertical markers indicate nuclear Bragg peak positions; lower red vertical markers in (b) indicate magnetic Bragg peak positions.

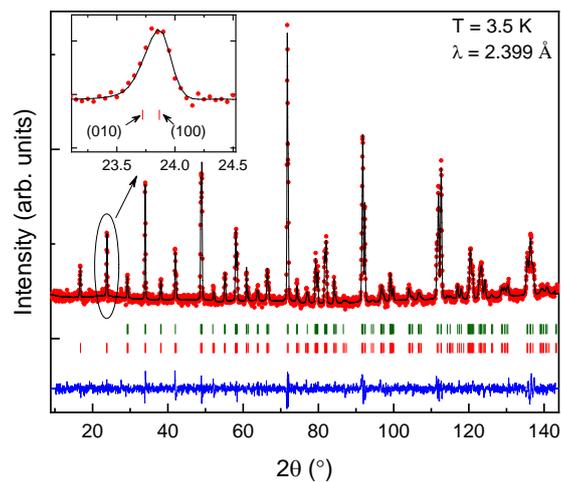

**Fig. 10:** Rietveld fit to NPD patterns collected at 3.5 K with λ = 2.399 Å on D2B diffractometer at ILL Red circles show observed data points; black line is calculated profile; lower blue line is difference profile (obs.-cal.); upper green vertical markers indicate nuclear Bragg peak positions; lower red vertical markers indicate magnetic Bragg peak positions. The inset enlarge the view near 24° to highlight the magnetic peak (010)/(100).

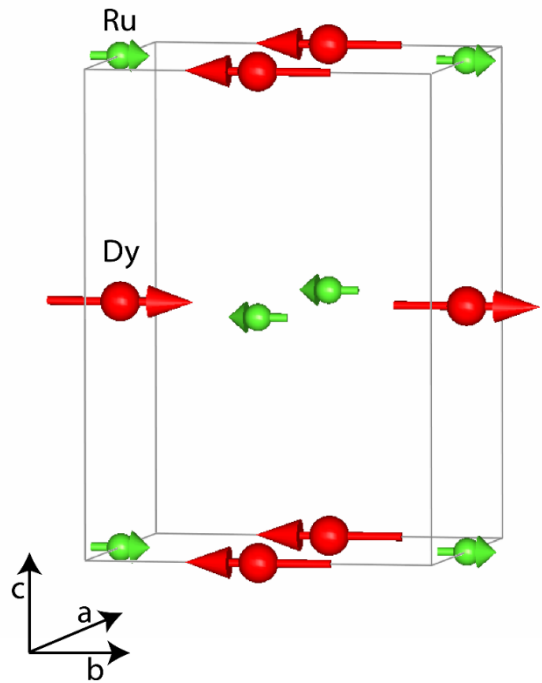

**Fig. 11:** The magnetic structure of Sr$_2$DyRuO$_6$ for $k = (0, 0, 0)$. The Dy$^{3+}$ and Ru$^{5+}$ moments are shown in green and red colors respectively.

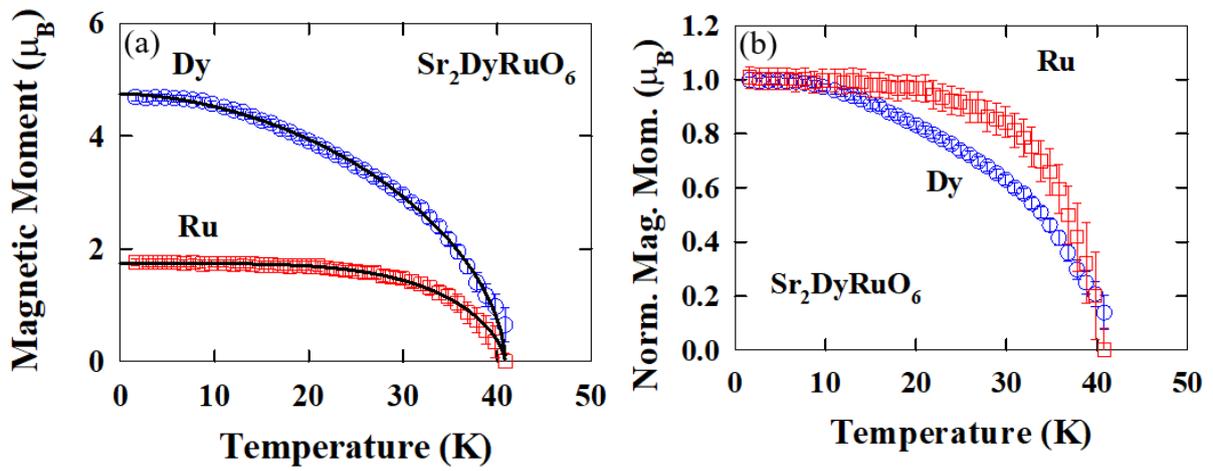

**Fig. 12:** Thermal variation of (a) Dy$^{3+}$ and Ru$^{5+}$ moments and (b) Normalized moments of Dy$^{3+}$ and Ru$^{5+}$.

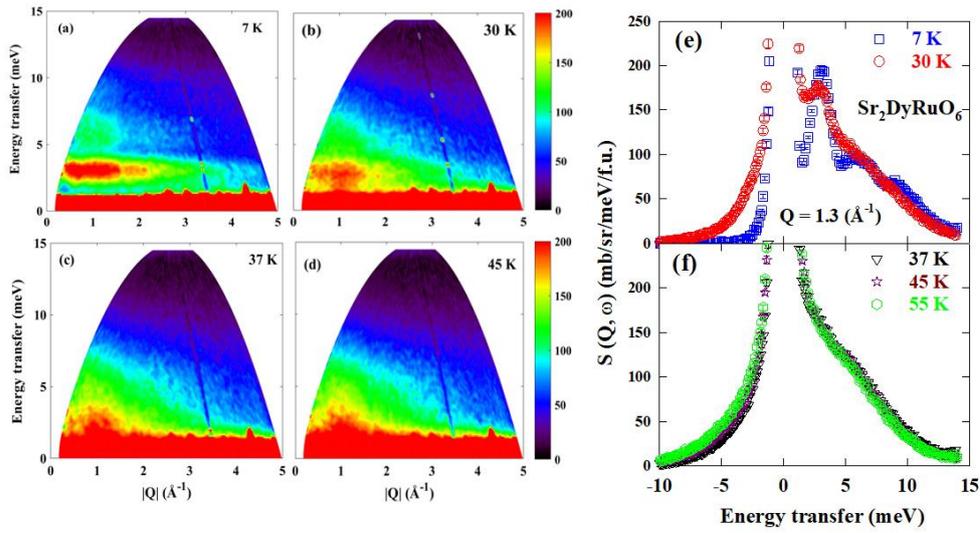

**Fig. 13:** (a-d) The color contour maps of scattering intensity versus momentum transfer at various temperature measure with an incident energy of $E_i$ = 15 meV at various temperatures on MERLIN. (e-f) The Q-integrated (Q=0 to 2 Å$^{-1}$) energy cuts at various temperatures between 7 and 55 K.

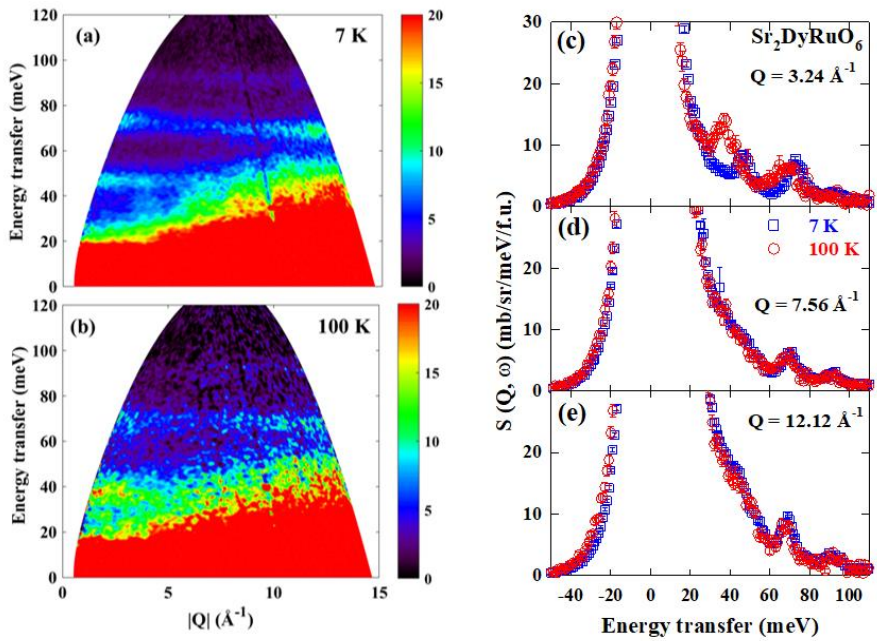

**Fig. 14:** (a-b) The color couture maps of scattering intensity versus momentum transfer at various temperature measure with an incident energy of $E_i$ = 130 meV at 7 and 100 K on MERLIN. (c-f) The Q-integrated energy cuts at low-Q, medium-Q and at high-Q at 7 and 100 K.

# SUPPLEMENTARY MATERIAL

## Muon spin rotation and neutron scattering investigations of the B-site ordered double perovskite $Sr_2DyRuO_6$

D.T. Adroja[1,2], Shivani Sharma[1,6], C. Ritter[3], A.D. Hillier[1], C.V. Tomy[4], R. Singh[5], R. I. Smith[1], M. Koza[3], A. Sundaresan[6], S. Langridge[1]

[1]*ISIS facility, Rutherford Appleton Laboratory, Chilton Oxon, OX11 0QX*

[2]*Highly Correlated Matter Research Group, Physics Department, University of Johannesburg, Auckland Park 2006, South Africa*

[3]*Institut Laue-Langevin, 71 Avenue des Martyrs, BP 156, F-38042 Grenoble Cedex, France*

[4]*Department of Physics, Indian Institute of Technology Bombay, Mumbai 400 076, India*

[5]*Indian Institutes of Science Education and Research, Bhopal, India*

[6]*Jawaharlal Nehru Centre for Advanced Scientific Research, Jakkur, Bangalore, India*

(Date: 29-07-2019)

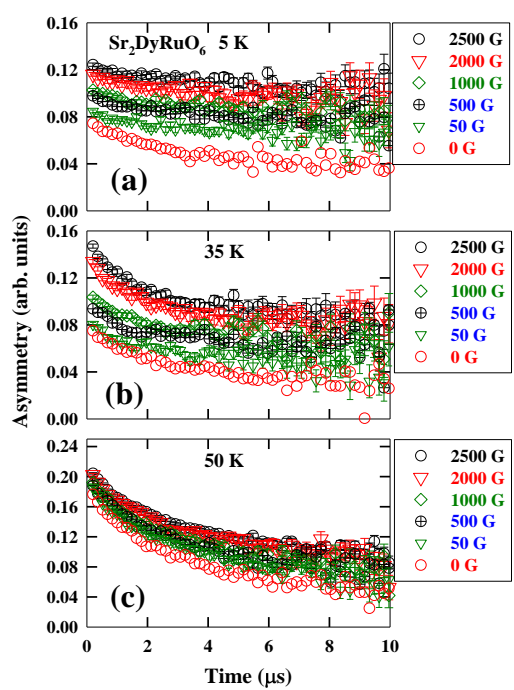

**Fig. S1** Longitudinal field (LF) µSR measurements at 5, 35 and 50 K.

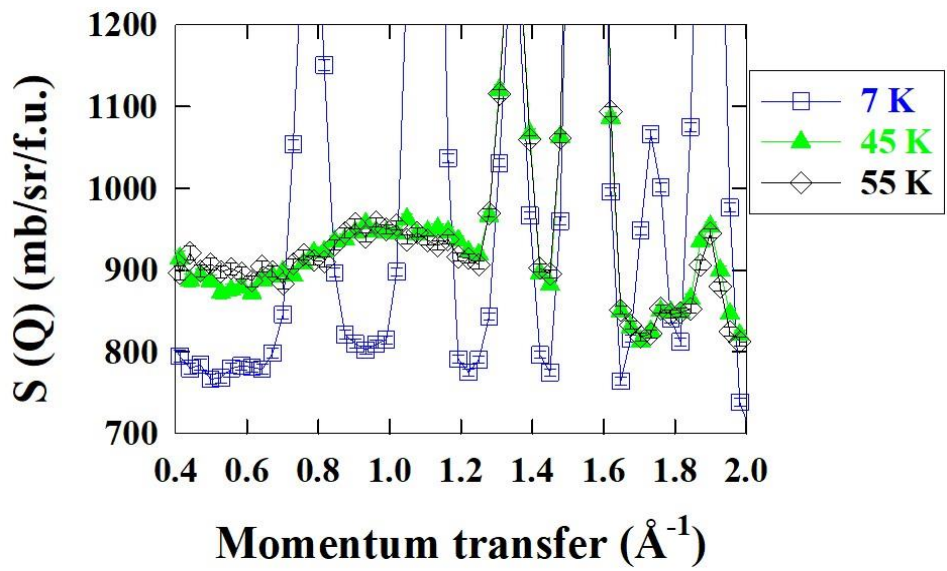

**Fig. S2** The energy integrated (-2 to 2 meV) Q versus intensity plot from MERLIN data of $E_i$ = 15 meV at 7 K, 45 K and 55 K. The extra peaks at 7 K are the magnetic Bragg peaks and at 45 K and 55 K very weak broad diffuse scattering near Q=1 Å$^{-1}$ can be seen.

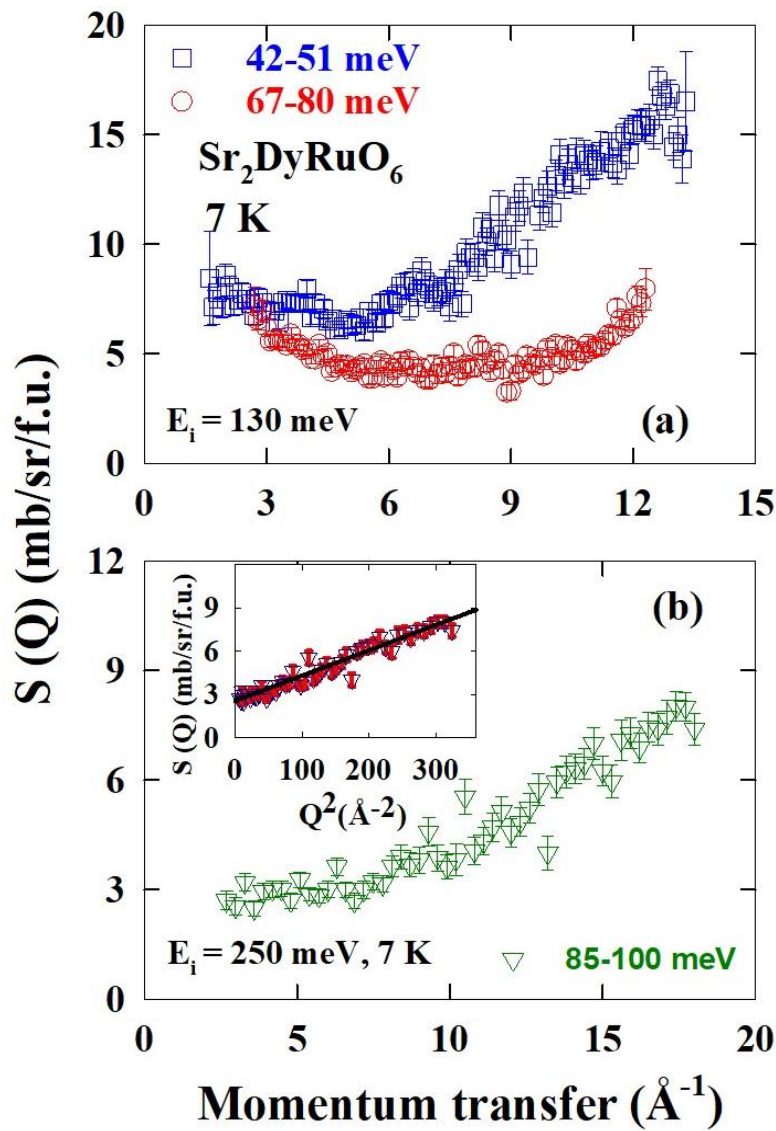

**Fig. S3.** Energy integrated, momentum (Q) dependence of scattering intensity of 46.8, 74.3 and 90.7 meV peaks at 7 K measured using $E_i$=130 meV (a) and 250 meV (b) on MERLIN. The inset in (b) shows the intensity vs $Q^2$ plot and the linear behaviour (solid line) reveals that this excitation is due to phonon model.

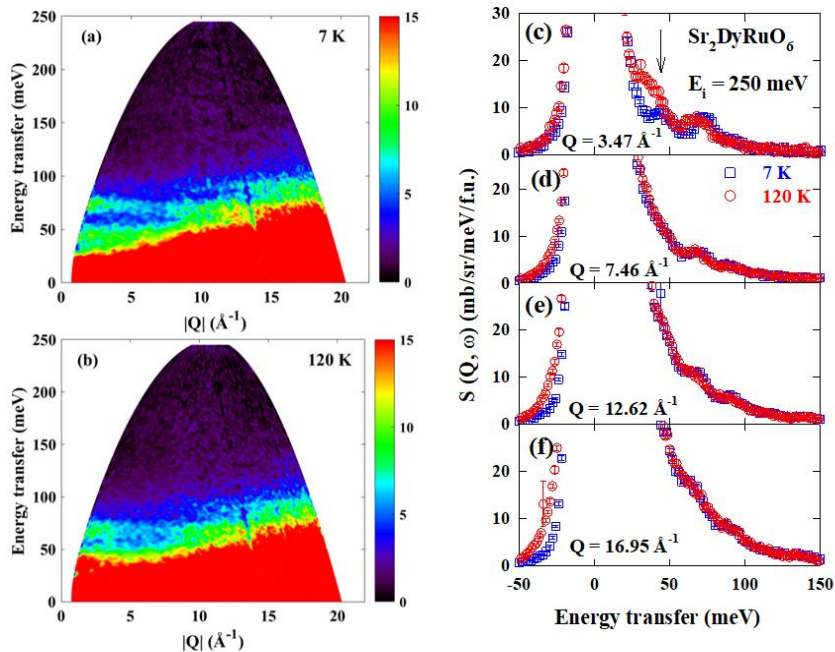

**Fig. S4 (a-b)** The colour couture maps of scattering intensity, energy transfer versus momentum transfer, at 5 K and 120 K measured with $E_i$=250 meV on MERLIN. (c-f) The Q-integrated energy cuts at Q= 0 to 5 Å$^{-1}$, 5 to 10 Å$^{-1}$, 10 to 15 Å$^{-1}$ and 15 to 20 Å$^{-1}$ at 7 K and 120 K. The arrow in Fig.(c) shows the excited state transition at 120 K.

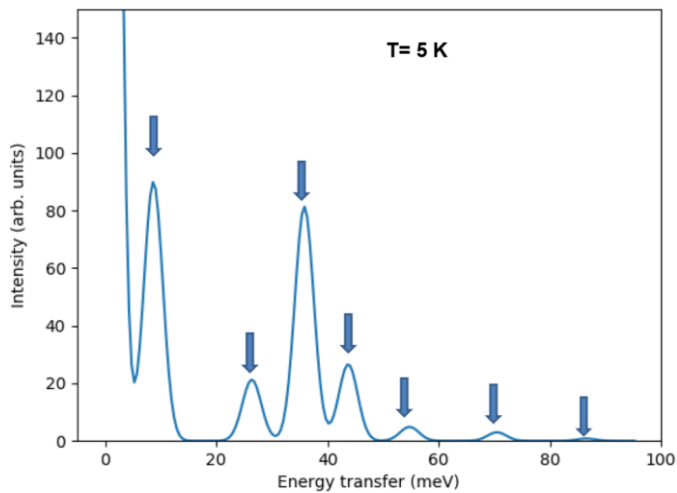

**Fig. S5** Crystal field excitations calculated using a point change model in Mantid-crystal field program [1] at 7 K. The linewidth of the excitations was take as 4 meV. Seven crystal field excitations (doubly degenerate) can be seen. The overall splitting qualitatively agrees with the experimental data.

1. O. Arnold, et al., Mantid—Data analysis and visualization package for neutron scattering and µSR experiments, *Nuclear Instruments and Methods in Physics Research Section A*, **764**, 156 (2014), http://dx.doi.org/10.1016/j.nima.2014.07.029.